\documentclass[aip,jcp,amsmath,reprint]{revtex4-1}

\usepackage{mathptmx}
\usepackage{afthd}
\usepackage{textcomp,graphicx,dcolumn}
\usepackage{hyperref}
\hypersetup{bookmarksnumbered=true,bookmarksopen,bookmarksopenlevel=1,
            colorlinks,allcolors=blue,urlcolor=black,
            pdfauthor={Vincent Holten, David T. Limmer, Valeria Molinero, and Mikhail A. Anisimov},
			bookmarksdepth=2,pdfpagemode=UseOutlines,
            pdftitle={Nature of the anomalies in the supercooled liquid state of the mW model of water},
            pdfprintscaling=None}

\usepackage[T1]{fontenc}
\usepackage[ansinew]{inputenc}

\newcommand{\ea}{\textit{et al.}}
\newcommand{\figref}[1]{Fig.~\ref{#1}}
\newcommand{\ffigref}[1]{Figure~\ref{#1}}

\renewcommand*{\eqref}[1]{Eq.~(\ref{#1})}

\newcommand*{\eqsref}[2]{Eqs.~(\ref{#1}) and (\ref{#2})}

\newcommand{\tabref}[1]{Table~\ref{#1}}

\newcommand{\pypxd}[3]{\left(\frac{\partial{#1}}{\partial{#2}}\right)_{#3}}

\newcommand*{\ts}[1]{_\text{#1}}
\newcommand*{\tu}[1]{^\text{#1}}

\newcommand*{\GnA}{G\tu{A}}
\newcommand*{\GnB}{G\tu{B}}
\newcommand*{\GnBA}{G\tu{BA}}
\newcommand*{\xe}{x\ts{e}}
\newcommand*{\Tc}{T\ts{c}}
\newcommand*{\Pc}{P\ts{c}}
\newcommand*{\Dt}{\Delta \hat{T}}
\newcommand*{\Ph}{\hat{P}}
\newcommand*{\Tx}{T\ts{i}}

\begin{document}

\title{Nature of the anomalies in the supercooled liquid state of the\\
mW model of water}
\author{Vincent Holten}
\affiliation{Institute for Physical Science and Technology and
  Department of Chemical and Biomolecular Engineering,\\
  University of Maryland, College Park, Maryland 20742, USA}
\author{David T. Limmer}
\affiliation{Department of Chemistry, University of California, Berkeley, California
94720, USA}
\author{Valeria Molinero}
\affiliation{Department of Chemistry, University of Utah, Salt Lake City, Utah
84112-0580, USA}
\author{Mikhail A. Anisimov}
\email[Author to whom correspondence should be addressed. Electronic mail: ]
{anisimov@umd.edu} \affiliation{Institute for Physical Science and Technology and
  Department of Chemical and Biomolecular Engineering,\\
  University of Maryland, College Park, Maryland 20742, USA}
\date{\today}

\begin{abstract}
The thermodynamic properties of the supercooled liquid state of the mW model of water
show anomalous behavior. Like in real water, the heat capacity and compressibility
sharply increase upon supercooling. One of the possible explanations of these anomalies,
the existence of a second (liquid--liquid) critical point, is not supported by
simulations for this model. In this work, we reproduce the anomalies of the mW model with
two thermodynamic scenarios: one based on a non-ideal ``mixture'' with two different
types of local order of the water molecules, and one based on weak crystallization
theory. We show that both descriptions accurately reproduce the model's basic
thermodynamic properties. However, the coupling constant required for the power laws
implied by weak crystallization theory is too large relative to the regular backgrounds,
contradicting assumptions of weak crystallization theory. Fluctuation corrections outside
the scope of this work would be necessary to fit the forms predicted by weak
crystallization theory. For the two-state approach, the direct computation of the
low-density fraction of molecules in the mW model is in agreement with the prediction of
the phenomenological equation of state. The non-ideality of the ``mixture'' of the two
states never becomes strong enough to cause liquid--liquid phase separation, also in
agreement with simulation results.
\end{abstract}

\maketitle

\section{Introduction}
The peculiar properties of supercooled water continue to gain interest. In the
supercooled region, the thermodynamic response functions, namely heat
capacity,\cite{angell1982} thermal expansivity,\cite{hare87,kanno1980} and
compressibility,\cite{kanno1979} show strong temperature dependences suggesting a
possible divergence at a temperature just below the homogeneous ice nucleation limit. One
of the scenarios to explain the anomalous behavior of real water is the existence of a
liquid--liquid transition terminated by the liquid--liquid critical
point.\cite{poole1992,mishima1998review,stanley2000,deben03,stokely2010,mishima2010review,%
mishima1998,mishima2000,mishima2011,muratatanaka2012} An explicit equation of state based
on this scenario is able to accurately represent all experimental data on the
thermodynamic properties of supercooled water.\cite{holtentwostate} If it exists, the
second critical point of water cannot be directly observed in a bulk experiment, because
it is located in ``no man's land,'' below the homogeneous ice nucleation
temperature.\cite{debenedettimetastablebook,kanno1975} Computer simulations of water can
provide additional insights into the nature of water's anomalies. The mW model devised by
Molinero and Moore\cite{molinero2009} represents the water molecule as a single atom with
only short-range interactions, and is is suitable for fast computations. The mW model
imitates the anomalous behavior of cold and supercooled water, including the density
maximum and the increase of the heat capacity and compressibility in the supercooled
region.\cite{molinero2009,moore2011,limmer2011} Using molecular dynamics simulations,
Limmer and Chandler\cite{limmer2011} have shown that the mW model does not exhibit a
second critical point or liquid--liquid separation in the range studied (0~MPa to
290~MPa, down to 170~K). Indeed, Moore and Molinero\cite{moore2011} have demonstrated
that in this model the supercooled liquid can no longer be equilibrated before it
crystallizes and there is no sign of a liquid--liquid transition at supercooling. This
raises the question: what is the origin of the anomalies in the mW model? In this work,
we explain and reproduce these anomalies with a thermodynamic equation of state based on
a non-ideal ``mixture'' of two kinds of molecular environments, in which the non-ideality
is mainly entropy driven. By analyzing experimental data with this equation of state,
Anisimov and coworkers\cite{bertrand2011,holtentwostate} have concluded that a
liquid--liquid transition can occur if effects of crystallization are neglected. We show
that for the mW model the nonideality of the free energy of mixing never becomes large
enough to cause liquid--liquid phase separation. Finally, we show that the properties of
the mW model may be described by power laws suggested by weak crystallization theory,
which predicts apparently diverging corrections to the regular thermodynamic properties
as a result of fluctuations in the translational order parameter, close to the limit of
stability of the liquid phase. However, the resulting value of the coupling constant is
large, which contradicts the basic assumption of the theory that the corrections must
always remain smaller than the regular backgrounds, suggesting that fluctuation
corrections beyond what are considered here may be important.

\section{Two-state thermodynamics of liquid water}
We assume liquid water at low temperatures to be a mixture of two interconvertible states
or structures, a high-density state A and a low-density state B. The fraction of
molecules in state B is denoted by $x$, and is controlled by the `reaction'
\begin{equation}
\text{A}\rightleftharpoons \text{B}.  \label{eq:reaction}
\end{equation}
The states A and B could correspond to different arrangements of the hydrogen-bonded
network.\cite{eisenbergbook_mixturemodels} In water-like atomistic models, such as the mW
model, these states could correspond to two kinds of local coordination of the water
molecules.\cite{moore2009}

In a real liquid, molecular configurations form a continuous distribution of coordination
numbers and local structures. Therefore, for real water the division of molecular
configurations into two states seems a gross simplification. However, such an approach
could serve as a first approximation if the distribution can be decomposed in two
populations with distinct properties. A similar concept, ``quasi-binary approximation,''
is commonly used to describe the properties of multi-component
fluids.\cite{valyashko2008} As for any phenomenological model, the utility of the
two-state approximation is to be provided by a comparison with experimental or
computational data.

Two-state equations of state have become popular to explain liquid polyamorphism.
\cite{mcmillan2004,mcmillan2006,vedamuthu1994,tanaka2012} Ponyatovsky
\ea\cite{ponyatovsky1998} and Moynihan\cite{moynihan1997} assumed that water could be
considered as a `regular binary solution' of two states, which implies that the phase
separation is driven by energy, and they qualitatively reproduced the thermodynamic
anomalies of water. Cuthbertson and Poole\cite{cuthbertsonpoole2011} showed that the ST2
water model can be described by a regular-solution two-state equation. Bertrand and
Anisimov\cite{bertrand2011} introduced a two-state equation of state where water is
assumed to be an athermal solution, which undergoes phase separation driven by non-ideal
entropy upon increase of the pressure. This equation of state was used by Holten and
Anisimov\cite{holtentwostate} to successfully describe the properties of real supercooled
water.

In general, the molar Gibbs energy of a two-state mixture is
\begin{equation}
    G = \GnA + x\GnBA + RT[x\ln x + (1-x)\ln(1-x)] + G\tu{E},
\end{equation}
where $R$ is the gas constant, $T$ is the temperature, $\GnBA\equiv \GnB-\GnA $ is the
difference in molar Gibbs energy between pure configurations A and B, and $G\tu{E}$ is
the excess Gibbs energy of mixing. The difference $\GnBA$ is related to the equilibrium
constant $K$ of ``reaction'' (\ref{eq:reaction}) as
\begin{equation}
\ln K(T,P)=\frac{\GnBA}{RT},
\end{equation}
where $P$ is the pressure. For the application to the mW model, we adopt a linear
expression for $\ln K$ as the simplest approximation.
\begin{equation}\label{eq:lnK}
\ln K=\lambda (\Dt+a\Ph),
\end{equation}
with
\begin{equation}\label{eq:dimensionlessTandP}
\Dt=(T-T_{0})/T_{0},\qquad \Ph=P/\rho _{0}RT_{0},
\end{equation}
where $T_{0}$ is the temperature at which $\ln K=0$ for zero pressure, and $\rho _{0}$ is
a reference density. The parameter $a$ is proportional to the slope of the $\ln K=0$ line
in the $T$--$P$ phase diagram. The parameter $\lambda $ is proportional to the heat of
reaction~(\ref{eq:reaction}), while the product $v=\lambda a$ is proportional to the
volume change of the reaction.

The excess Gibbs energy,
\begin{equation}
G\tu{E} = H\tu{E}-T S\tu{E},
\end{equation}
causes the non-ideality of the mixture and is the sum of contributions of
the enthalpy of mixing $H\tu{E}$ and excess entropy $S\tu{E}$.

The Gibbs energy $\GnA$ of the pure structure A defines the background of
the properties and is approximated as
\begin{equation}  \label{eq:background}
\GnA = RT_0 \sum_{m,n} c_{mn}(\Dt)^m \Ph^n,
\end{equation}
where $m$ and $n$ are integers and $c_{mn}$ are adjustable coefficients.

\begin{figure*}[tbp]
\includegraphics{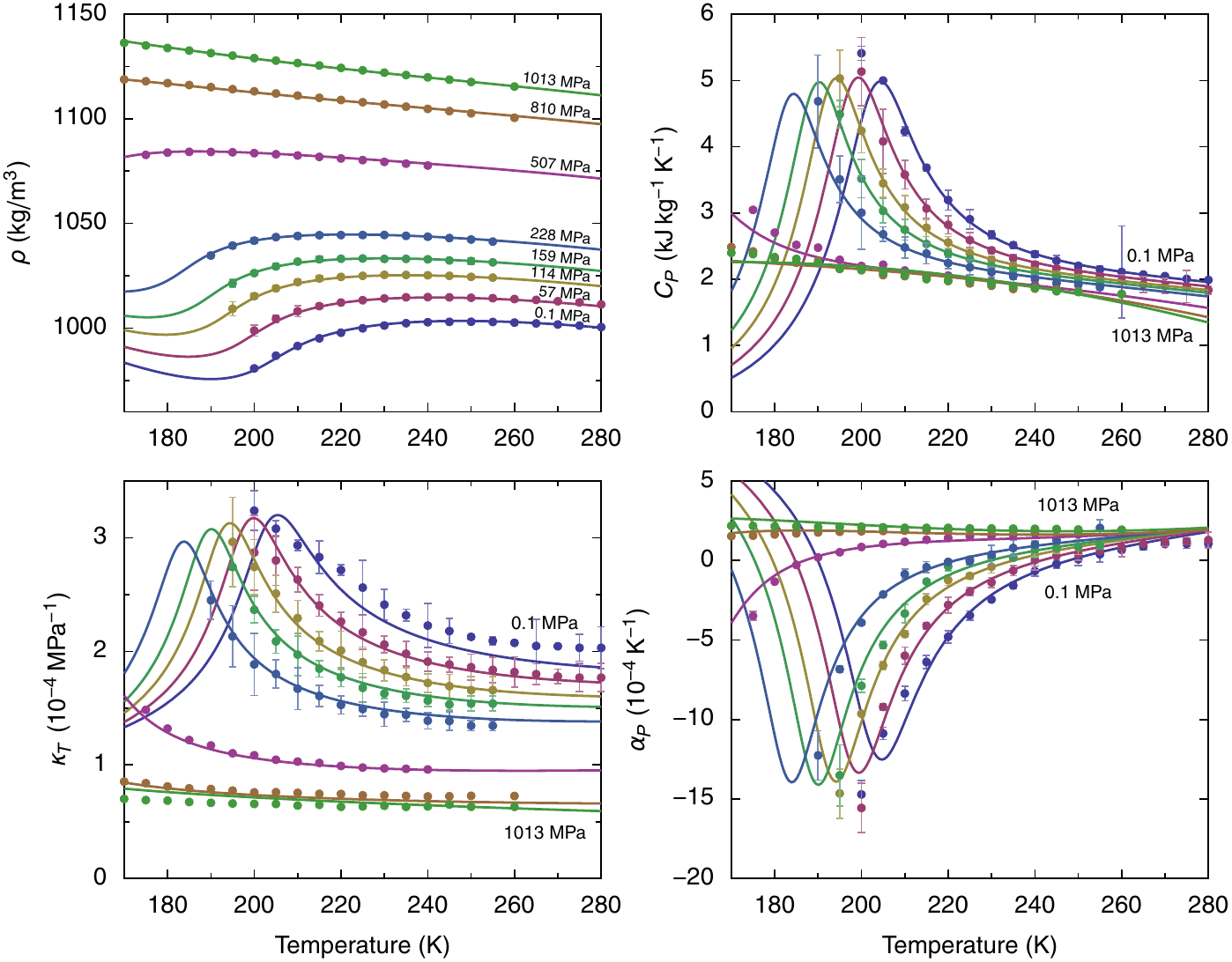}
\caption{\label{fig:TS}Density $\rho$, isobaric heat capacity $C_P$, isothermal
compressibility $\kappa_T$, and thermal expansivity $\alpha_P$ computed from the mW model
(points) compared with the results (curves) of the two-state approach for an athermal
solution [\eqref{eq:athermalsolution}]. The isobar pressures are given in the density
diagram; pressures and corresponding isobar colors are the same for the other plots.}
\end{figure*}

\subsection*{Regular solution}
In the case of a regular solution, the non-ideality is entirely associated with the
enthalpy of mixing. Taking $H\tu{E} = w x(1-x)$, we obtain
\begin{equation}  \label{eq:regularsolution}
G = \GnA + x\GnBA + RT[x\ln x + (1-x)\ln(1-x)] + w x(1-x).
\end{equation}
The interaction parameter $w$ is assumed to be independent of temperature,
but may depend on pressure. Considering $x$, the fraction of B, as the
reaction coordinate or extent of reaction, \cite{prigogine1954} the
condition of chemical reaction equilibrium,
\begin{equation}
\left( \frac{\partial G}{\partial {x}}\right) _{T,P}=0,
\end{equation}
defines the equilibrium fraction $\xe$ through
\begin{equation}  \label{eq:xeregular}
\ln K+\ln \frac{\xe}{1-\xe}+\frac{w}{RT} (1-2\xe) =0.
\end{equation}
If the excess Gibbs energy becomes large enough, phase separation may occur,
and a critical point may exist in the phase diagram. In the case of a
regular solution, \eqref{eq:regularsolution}, the conditions for the
critical point of liquid--liquid equilibrium,
\begin{equation}  \label{eq:criticalpointconditions}
\left(\frac{\partial^2 G}{\partial x^2}\right)_{T,P} = 0, \qquad
\biggl(\frac{\partial^3 G}{\partial x^3}\biggr)_{T,P}=0,
\end{equation}
yield the critical composition $x\ts{c}=1/2$ and the condition
\begin{equation}  \label{eq:regularcriticalomega}
T = \frac{w(P)}{2R}.
\end{equation}
If the line in the phase diagram given by \eqref{eq:regularcriticalomega} intersects the
line given by the phase equilibrium condition $\ln K(T,P) = 0$, a critical point exists
at the intersection and a line of liquid--liquid phase equilibrium emanates from this
point toward lower temperature. This is an energy-driven phase transition, like in the
lattice-gas model.

\subsection*{Athermal solution}
In the case of an athermal solution, the enthalpy of mixing is zero and the non-ideality
is associated with the excess entropy of mixing. With the simplest symmetric form of
excess entropy, $S\tu{E} = -R \omega x(1-x)$, we obtain
\begin{equation}  \label{eq:athermalsolution}
G = \GnA + x\GnBA + RT[x\ln x + (1-x)\ln(1-x) + \omega x(1-x)].
\end{equation}
The equilibrium fraction $\xe$ is given by
\begin{equation}  \label{eq:xeathermal}
\ln K+\ln \frac{\xe}{1-\xe}+\omega (1-2\xe)=0.
\end{equation}
The critical-point conditions of \eqref{eq:criticalpointconditions} yield
the critical value
\begin{equation}
\omega\ts{c} = 2.
\end{equation}
Thus, the critical pressure $\Pc$ is the pressure at which the function $\omega(P)$
reaches the value 2. The critical temperature $\Tc$ follows from the phase-equilibrium
condition, $\ln K(\Tc,\Pc) = 0$. If the function $\omega(P)$ remains below 2 for all
pressures, a critical point does not exist and the mixture does not phase separate. If
$\omega$ is larger than 2 for a certain pressure, an entropy-driven phase transition
exists. While real water shows a preference for the athermal two-state
description,\cite{holtentwostate} the experimental data cannot exclude a contribution of
energy in the non-ideal part of the Gibbs energy.

\subsection*{Clustering of water molecules}
Moore and Molinero\cite{moore2009} have demonstrated that four-coordinated molecules
cluster in supercooled mW water, an idea originally proposed by Stanley and
Teixeira.\cite{stanley1980} The formation of clusters of molecules belonging to a single
state reduces the number of configurations and decreases the mixing entropy. Clustering
can be incorporated in the equation of state by dividing the ideal mixing entropy terms
by the number of molecules in a cluster $N$. For the athermal solution,
\eqref{eq:athermalsolution}, this yields
\begin{align}  \label{eq:athermalsolutionwithclustering}
    G =~&\GnA + x\GnBA\notag\\
     &+ RT\left[\frac{x}{N_1}\ln x + \frac{1-x}{N_2}\ln(1-x) + \omega x(1-x)\right].
\end{align}
Formally, this expression is identical to the free energy of a mixture of two polymers
with (large) degrees of polymerization $N_1$ and $N_2$ in Flory--Huggins
theory.\cite{florybook} However, in our case these parameters are purely
phenomenological. In principle, the cluster sizes $N_1$ and $N_2$ are functions of
temperature and pressure. In this work, we use a constant $N_1 = N_2 = N$ as a first
approximation. Furthermore, one can imagine a mixed scenario in which the system combines
both athermal-solution and regular-solution features.

\subsection*{Description of thermodynamic properties of the mW model}

\begin{figure}
\includegraphics{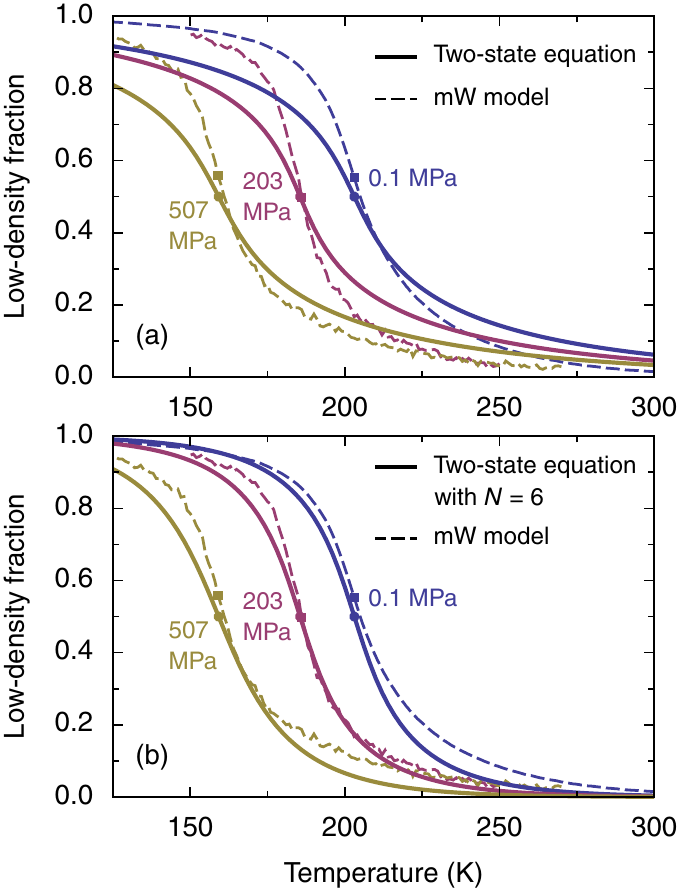}
\caption{\label{fig:fractionrenorm}Fraction $x$ of molecules in the low-density state,
Solid curves: fraction $x$ for the
(a) two-state equation of state, \eqref{eq:athermalsolution};
(b) two-state equation with account for hexamer clustering, \eqref{eq:athermalsolutionwithclustering} with $N=6$.
Dashed curves: fraction $x$ obtained from simulations of mW water,
calculated from the fraction of four-coordinated molecules $f_4$ as
$x = (f_4 - f_4\tu{H})/(f_4\tu{L} - f_4\tu{H})$, to account for fractions $f_4\tu{L}$ and $f_4\tu{H}$ of
four-coordinated molecules in the low- and high-temperature liquid, respectively.
The inflection points on the curves are marked with circles (two-state equation) and
squares (mW model).
The data was collected by linearly quenching the temperature of the simulations at a rate of 10 K/ns.
The data below the inflection point do not correspond to equilibrium states.}
\end{figure}

\begin{figure}
\includegraphics{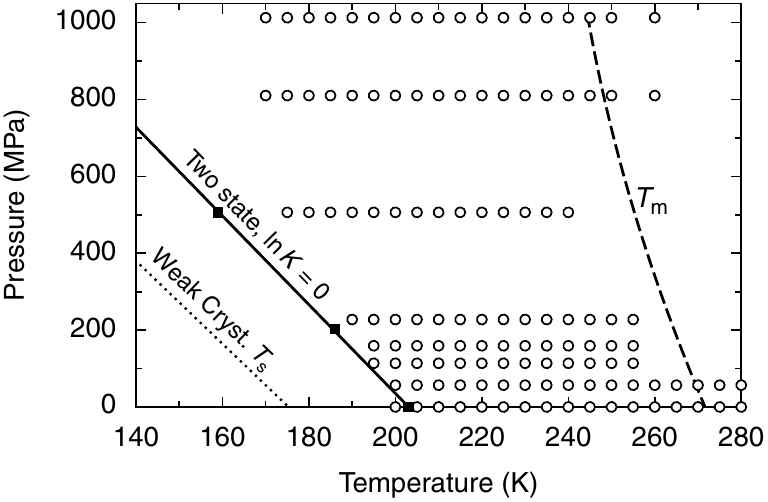}
\caption{\label{fig:PTdiagram}Pressure--temperature diagram.
Circles: location of the computed property data of the mW model.
Solid line: line at which $\ln K = 0$ and the low-density fraction $x=1/2$ for the two-state equation.
Squares: location of inflection points of the low-density fraction $x$ for the mW model
(see \figref{fig:fractionrenorm}).
Dotted line: stability-limit temperature $T\ts{s}$ from the fit of the weak crystallization
model to the mW data, \eqref{eq:Tsfit}.
The dashed curve is a fit to the melting temperature $T\ts{m}$ of mW ice (uncertainty about $\pm$ 3~K),
obtained from free energy calculations as described in Ref.~\onlinecite{limmer2011}.}
\end{figure}

Thermodynamic properties of the mW model, namely density, isothermal compressibility, and
heat capacity have been calculated from molecular dynamics simulations by Limmer and
Chandler up to 228~MPa.\cite{limmer2011} For this work, the expansion coefficient has
also been calculated, and the calculations have been extended to higher pressures, as
shown in \figref{fig:TS}. We apply the two-state thermodynamics of water to explain and
reproduce the calculated properties and demonstrate that this model does not show
liquid--liquid separation, at least in the range of pressures and temperatures studied.

First, we have verified whether an ``ideal-solution'' two-state thermodynamics, for which
the excess Gibbs energy is zero, can reproduce the thermodynamic properties of the mW
model. Indeed, the competition between these two states, even without non-ideality, can
cause the density maximum and the increase in the response functions. However, the
thermodynamic properties are not the only data that should be matched. Moore and
Molinero\cite{moore2009} have calculated the fraction of four-coordinated molecules
$f_{4}$, which is a fraction of molecules in a low-density state
(Fig.~\ref{fig:fractionrenorm}). Specifically, $f_{4}$ is the fraction of molecules with
four neighbors within the first coordination shell up to a certain cutoff radius
$r\ts{c}$. The exact value of $f_{4}$ depends on the value of $r\ts{c}$, but the
inflection point of the $f_{4}(T)$ curve is independent of $r\ts{c}$ and occurs at
$\Tx=(201\pm 2)$~K at atmospheric pressure.\cite{moore2009} The value of $r\ts{c}$ was
0.35~nm for all pressures in this study. This cutoff corresponds to the first solvation
shell, as the position of the first minimum of the radial distribution function of mW
water is 0.35~nm at 0.1~MPa and 0.342~nm at 507~MPa. The fraction $x$ of low-density
liquid is related to $f_4$ by
\begin{equation}
    x(T) = \frac{f_4(T) - f_4\tu{H}}{f_4\tu{L} - f_4\tu{H}},
\end{equation}
which accounts for the finite fraction $f_4\tu{H}$ of four-coordinated molecules in the
high-temperature liquid, and the fraction $f_4\tu{L} < 1$ of four-coordinated molecules
in the low-temperature liquid. Both $f_4\tu{H}$ and $f_4\tu{L}$ are estimated by an
extrapolation of the fraction $f_4$ to high and low temperature. Below $\Tx$, liquid mW
water cannot be equilibrated without crystallization.\cite{moore2011} Nevertheless, the
fraction $f_4$ was also computed below $\Tx$ as in Ref.~\onlinecite{moore2009}: in
quenching simulations the temperature was varied linearly at 10~K/ns, the slowest rate
that results in the vitrification of mW water. Liquid mW water can be equilibrated down
to $\Tx$ at a cooling rate of 10~K/ns, but any property extracted from the quenching
simulations at $T < \Tx$ may depend on the cooling rate.

In the two-state thermodynamics, as given by \eqref{eq:xeregular} or
\eqref{eq:xeathermal}, the inflection point of the fraction $x$ at atmospheric pressure
occurs where $\ln K = 0$, near the temperature $T_{0}$ [\eqref{eq:dimensionlessTandP}].
To match the low-density fraction of the mW model, $T_{0}$ should be close to $\Tx$. This
is not the case for the ideal-solution two-state version, where $T_{0}=(160\pm 4)$~K,
significantly below~$\Tx$.

With a nonzero excess Gibbs energy, the two-state approach is able to reproduce both the
thermodynamic properties and the inflection point of the fraction in the mW model. As
\figref{fig:PTdiagram} shows, the inflection points of the fraction of low-density liquid
in mW simulations at different pressures form a straight line in the phase diagram. Since
the two-state equation yields inflection points at the line $\ln K = 0$, the inflection
points of the mW model can be matched by adopting suitable values of the slope $a$ and
intercept $T_0$ of the $\ln K = 0$ line [\eqref{eq:lnK}]. When the $\ln K = 0$ line is
fixed in this way, the athermal-solution (entropy-driven non-ideality) version yields a
better description than the regular-solution (energy-driven non-ideality) version; the
sum of squared deviations of the regular-solution fit from the thermodynamic property
data is about 50\% higher than that of the athermal-solution fit. More convincing
evidence in favor of the athermal-solution approximation comes from the direct
computation of the enthalpy and entropy of mW water; see below. An approximation with a
constant interaction parameter, $w$ or $\omega$, works reasonably well, and the
description can be further improved by making the interaction parameter weakly dependent
on pressure. In the case of the athermal-solution version, the two-state thermodynamics
for the mW model excludes liquid--liquid separation at any temperature or pressure,
because the value of $\omega$ is lower than 2 (for a pressure-independent $\omega$, the
optimum value is 1.61). In the case of the regular-solution version, the two-state
thermodynamics for the mW model predicts liquid--liquid separation at a temperature below
147~K, which is outside the kinetic limit of metastability of mW liquid
water.\cite{moore2011} \ffigref{fig:TS} shows the predicted thermodynamic properties for
the athermal-solution case with a quadratic pressure dependence of $\omega$, and
Fig.~\ref{fig:fractionrenorm}a shows the low-density fraction. The numerical values of
all parameters are given in the appendix. At 0.1~MPa, there is a systematic difference
between compressibility values calculated from the two-state equation and those of the mW
model. This difference can be decreased by including more background terms, but that
would make the description more empirical.

An improved description of the low-density fraction is obtained when clustering of water
molecules is taken into account in the equation of state, as in
\eqref{eq:athermalsolutionwithclustering}. When the number of molecules $N$ in a cluster
is taken as an adjustable parameter in the fit, the optimum value is 6.5, but the quality
of the fit varies little for values of $N$ between 4 and 10. For clusters of $N=6$
molecules (hexa\-mers), the fraction that results from the fit to the mW properties is
shown in \figref{fig:fractionrenorm}b. Because of the division of the ideal mixing
entropy terms by $N$ in \eqref{eq:athermalsolutionwithclustering}, a smaller non-ideality
term is sufficient. Indeed, for the fit with $N=6$, the value of the interaction
parameter $\omega$ is 0.2, an order of magnitude smaller than in the case without cluster
formation. For such a small value of $\omega$, the difference between a regular and an
athermal solution becomes unimportant, because the main contribution to the nonideality
comes from clustering and thus is entropy driven. The work of Moore and
Molinero\cite{moore2009} shows that clustering of four-coordinated molecules increases
with cooling, which suggests that the parameter $N_1$ is temperature dependent. Such a
temperature-dependent $N_1$ would affect the values of the properties calculated with the
two-state thermodynamics, particularly at low temperature.

\begin{figure}
\includegraphics[width=8.3cm]{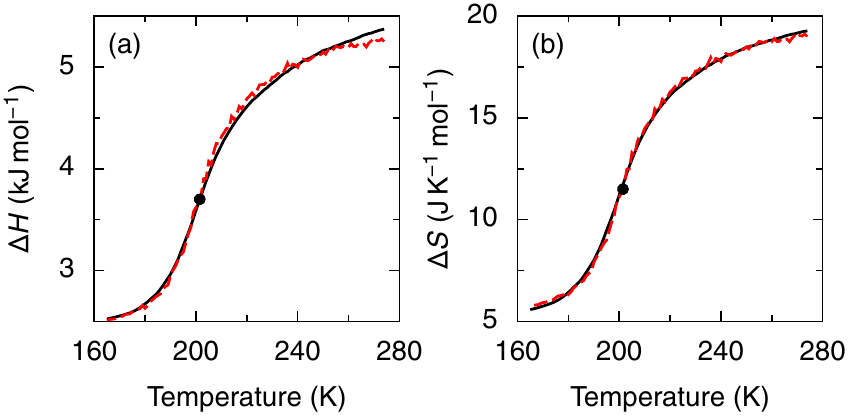}
\caption{\label{fig:HandS}
Enthalpy $\Delta H$ (a) and entropy $\Delta S$ (b) of liquid mW water with respect to ice at 0.1~MPa
(black curves, from Moore and Molinero\cite{moore2011})
and their fits (dashed) according to \eqsref{eq:Hfit}{eq:Sfit}, respectively.
Both $\Delta H$ and $\Delta S$ are computed at a cooling rate of 10~K/ns, which prevents
crystallization, so that the values below 200~K do not correspond to an equilibrium state.
The circle signals $\Tx = 201$~K.
These results support the modeling of mW water as an athermal mixture of two states.
}
\end{figure}

Analysis of the enthalpy of liquid water supports the conjecture that the excess free
energy of mixing is almost entirely due to entropic effects. If water is represented as a
``mixture,'' then its enthalpy can be written in terms of the partial molar enthalpies of
the low and high-density ``components''. \ffigref{fig:HandS} shows that the enthalpy of
liquid water with respect to ice in mW water is very well represented by a sum of the
weighted contributions of the two pure components, i.e., no excess enthalpy of mixing:
\begin{equation}\label{eq:Hfit}
   \Delta H = H\ts{liq} - H\ts{ice} = [1.84 x + 5.38 (1-x)]~\text{kJ/mol}.
\end{equation}
Here we use the enthalpy with respect to ice, instead of the enthalpy itself, to
eliminate the trivial temperature dependence of the partial molar enthalpies of the
components. This is possible in mW water because the heat capacity of the
four-coordinated component is almost indistinguishable from that of mW ice, and the heat
capacity of the high-temperature component is also quite close to that of ice, as a
result of the monatomic nature of the model. Thus, the temperature dependence of the
molar enthalpy difference between a component and ice is negligible.

The enthalpy of the low-density component with respect to ice, 1.84~kJ/mol, is in good
agreement with the $(1.35 \pm 0.15)$~kJ/mol measured for LDA in
experiments.\cite{floriano1989} The enthalpy of the high-density component with respect
to ice is, unsurprisingly, the enthalpy of melting of ice (5.3 kJ/mol for mW, 6.0 kJ/mol
in experiments). These results support the interpretation that there is no excess
enthalpy of mixing contribution to the non-ideal excess free energy of liquid mW water.

Interestingly, the entropy of liquid water with respect to ice can also be represented as
a weighted sum of temperature-independent contributions from the two pure components
(\figref{fig:HandS}):
\begin{equation}\label{eq:Sfit}
    \Delta S = S\ts{liq} - S\ts{ice} = [2.18 x + 19.75 (1-x)]~\mathrm{J/(K\,mol)},
\end{equation}
where the entropy of the low-density component with respect to ice, 2.18~J/(K\,mol), is
in good agreement with the experimental value for LDA, $(1.6 \pm
1.0)$~J/(K\,mol),\cite{smith2011} and the entropy of the high-density component with
respect to ice is the entropy of melting (19.3 J/(K\,mol) for mW water, 22 J/(K\,mol) in
experiment). The implication of \eqref{eq:Sfit} is that mW water is an ``athermal
solution,'' the excess (non-ideal) entropy of mixing is negative, and -- remarkably~-- it
essentially cancels out the positive ideal entropy of mixing contribution. This
cancellation strongly supports the idea of molecular clustering and is similar to the
thermodynamics of near-athermal mixtures of two high-molecular-weight polymers, where the
entropy of mixing is very small. This property of mW water excludes the possibility of
the regular-solution approximation.

\begin{figure}
\includegraphics{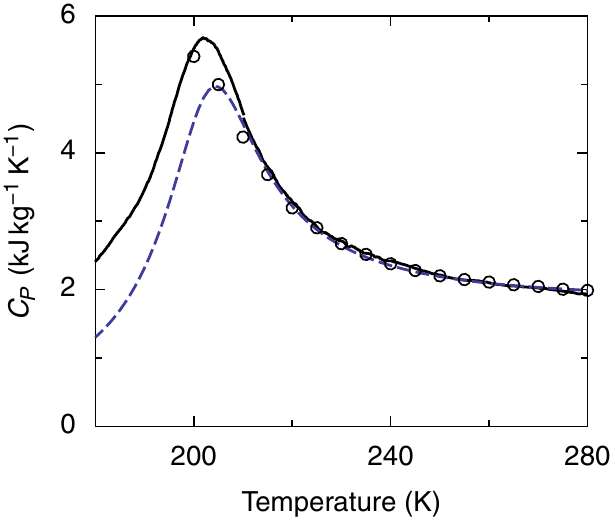}
\caption{\label{fig:CP}
Heat capacity of mW water in equilibrium (circles) and on hyperquenching at
10~K/ns (black curve, computations by Moore and Molinero\cite{moore2011}).
The values below 200~K do not correspond to an equilibrium state.
The dashed curve is the prediction of the two-state equation with hexamer clusters.}
\end{figure}

\ffigref{fig:CP} shows the isobaric heat capacity $C_P$ calculated for liquid mW water
down to a temperature below the crystallization temperature, which is made possible by
hyperquenching at 10~K/ns. Below 200~K, the two-state thermodynamics predicts a $C_P$
that is lower than the value for hyperquenched mW water. This discrepancy may be caused
by an oversimplicity of our equation of state, in particular the symmetric form of the
excess entropy of mixing $S\tu{E} = -R \omega x(1-x)$ and the constant value of $N$.

The two-state thermodynamics predicts maxima of the heat capacity and compressibility,
and minima of the density and expansivity near the line $\ln K = 0$. These extrema are
not observed in simulations of the equilibrium mW liquid, shown in \figref{fig:TS}, but
they are seen when the system is driven out of equilibrium through fast cooling rates, as
in \figref{fig:CP}. Furthermore, the two-state thermodynamics cannot predict the
stability limit of the liquid phase or account for any pre-crystallization effects.

\newcommand*{\Ts}{T_\text{s}}
\newcommand*{\Th}{\hat{T}}
\newcommand*{\Ths}{\hat{T}_\text{s}}
\newcommand*{\Tsz}{T_\text{s0}}
\newcommand*{\Vh}{\hat{V}}
\newcommand*{\Vc}{V_\text{c}}
\newcommand*{\mh}{\hat{\mu}}
\newcommand*{\kap}{\hat{\kappa}_T}
\newcommand*{\alp}{\hat{\alpha}_P}
\newcommand*{\Sh}{\hat{S}}
\newcommand*{\Cph}{\hat{C}_P}
\newcommand*{\Cvh}{\hat{C}_V}
\newcommand*{\mr}{\hat{\mu}^\text{r}}
\newcommand*{\amp}{\smash{\hat{\beta}}}

\section{Weak crystallization theory}
\begin{figure}
\includegraphics{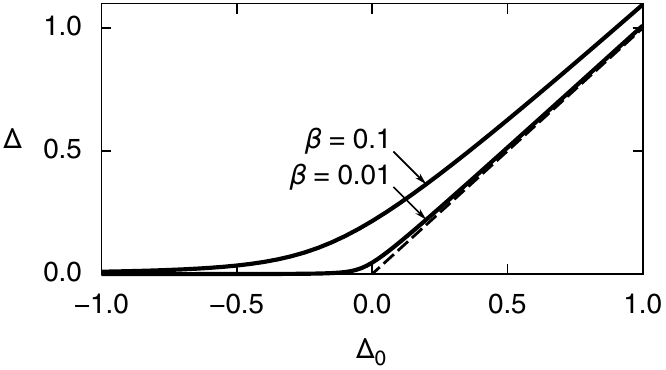}
\caption{\label{fig:delta}
Fluctuation-renormalized distance to the stability-limit temperature $\Delta$, given by \eqref{eq:gap},
as a function of the mean-field distance to the stability limit $\Delta_0$,
for two values of $\beta$ (solid curves). The dashed line corresponds to
$\Delta = \Delta_0$ and is shown as a reference.}
\end{figure}

According to the theory of weak
crystallization,\cite{brazovskii1975,brazovskii1987,kats1993} fluctuations of the
translational order parameter cause corrections in the thermodynamic response functions
close to the absolute stability limit of the liquid phase with respect to
crystallization. The distance to the stability limit $\Delta$ is defined as
\begin{equation}
    \Delta(T,P) = \frac{T - \Ts(P)}{\Ts(P)},
\end{equation}
where $T$ is the temperature, $P$ is the pressure, and $\Ts(P)$ is the stability-limit
temperature. According to this theory, the fluctuations of the translational
(short-wavelength) order parameter $\psi$ renormalize the mean-field distance $\Delta_0 =
(T - T\ts{s}\tu{MF})/T\ts{s}\tu{MF}$ between the temperature $T$ and the mean-field
absolute stability limit $T\ts{s}\tu{MF}$ of the liquid phase:
\begin{equation}\label{eq:gap}
    \Delta = \Delta_0 + \beta\Delta^{-1/2},
\end{equation}
where $\Delta$ is the fluctuation-renormalized distance to the stability-limit
temperature, and $\beta$ is a molecular parameter, similar to the Ginzburg number that
defines the validity of the mean-field approximation in the theory of critical phenomena.
Solutions of \eqref{eq:gap} are shown in \figref{fig:delta}. Fluctuations of the
translational order parameter stabilize the liquid phase, shifting the stability limit of
the liquid below the mean-field value, meaning that $\Delta_0$ becomes negative at
$\Delta$ tending to zero. This also means that the fluctuation part would not actually
diverge. Thus the theory of weak crystallization requires $\beta\Delta^{-1/2} \ll
\Delta_0$ and $\Delta_0 \ll 1$. As seen in \figref{fig:delta}, even the value of the
coupling constant $\beta = 0.1$ is already beyond the validity limit of the theory since
it corresponds to the mean-field gap $\Delta_0 = -1$.

\begin{figure*}
\includegraphics{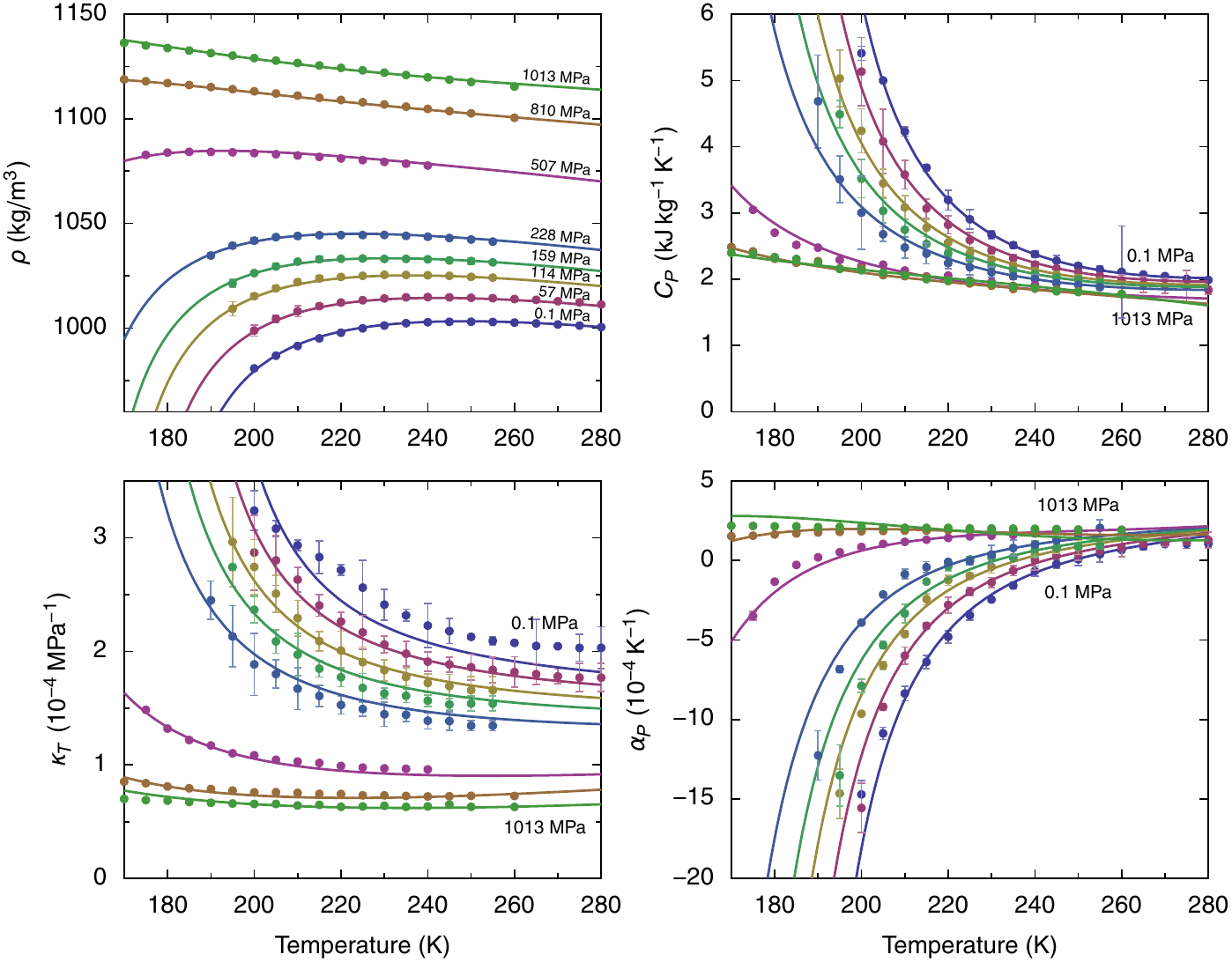}
\caption{\label{fig:WC}Density $\rho$, isobaric heat capacity $C_P$, isothermal
compressibility $\kappa_T$, and thermal expansivity $\alpha_P$ computed from the mW model
(points) compared with the fit to power laws given by weak crystallization theory
(curves). The isobar pressures are given in the density diagram; pressures and
corresponding isobar colors are the same for the other plots.}
\end{figure*}

The contribution of order-parameter fluctuations to the isobaric heat capacity $C_P$ and
isothermal compressibility $\kappa_T$ is proportional to $\Delta^{-3/2}$, while the
contribution to the density and entropy is ${\sim}\Delta^{-1/2}$. Accordingly, the
fluctuation contribution $\delta\mh = \amp\Delta^{1/2}$ can be included in the chemical
potential $\mh$ as
\begin{equation}\label{eq:gibbsWC}
    \mh = \amp\Delta^{1/2} + \mr(\Th,\Ph),
\end{equation}
where $\amp\simeq 2\beta$ (see the appendix) is the amplitude of the fluctuation part,
and $\mr$ is the regular (background) part of the chemical potential. The contributions
from fluctuations are to be small compared to the background. The variables with a hat in
\eqref{eq:gibbsWC} are dimensionless variables, defined as
\begin{equation}
    \Th=\frac{T}{\Tsz}, \quad
    \mh=\frac{\mu}{R \Tsz}, \quad
    \Ph=\frac{P V_0}{R \Tsz},
\end{equation}
where $\Tsz$ is the stability-limit temperature at zero pressure, $R$ is the molar gas
constant, and $V_0 = M\times 10^{-3}$~m$^3/$kg is an arbitrary reference constant for the
molar volume, with $M$ the molar mass of water. For our application of the weak
crystallization theory to the mW model results, we represent the regular part $\mr$ by a
truncated Taylor-series expansion,
\begin{equation}\label{eq:backgr}
    \mr(\Th,\Ph) = \sum_{m,\,n} c_{mn} \Th^m \Ph^n,
\end{equation}
similar to \eqref{eq:background}. We approximate the dependence of the stability-limit
temperature on the pressure by a linear function,
\begin{equation}\label{eq:Tsline}
    \Ths(\Ph) = 1 - a' \Ph,
\end{equation}
where the slope $a'$ is an adjustable parameter. Expressions for the thermodynamic
properties, found from derivatives of the chemical potential of \eqref{eq:gibbsWC}, are
given in the appendix.

\subsection*{Fit to the mW data}
\begin{figure}
\includegraphics{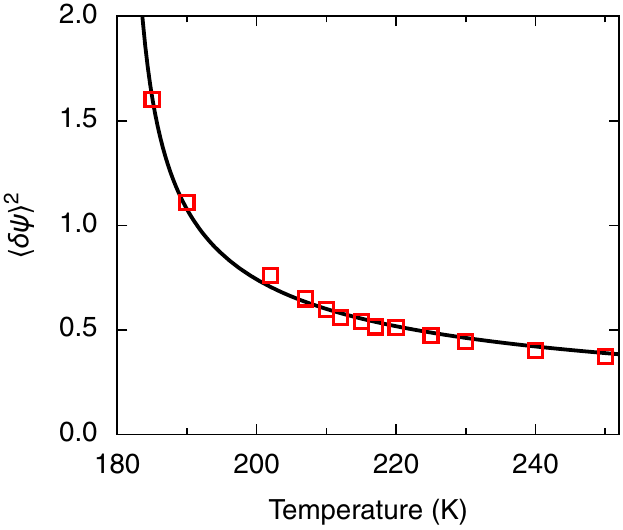}
\caption{\label{fig:qoft}Fluctuations of the crystallization order parameter from weak crystallization theory
(i.e. short-wavelength density fluctuations) in arbitrary units
as a function of temperature, fit with \eqref{eq:orderparameterfit} (curve).
}
\end{figure}

A fit of \eqref{eq:gibbsWC} to the mW data, shown in \figref{fig:WC}, results in a
stability-limit temperature of
\begin{equation}\label{eq:Tsfit}
    T\ts{s}(P)/K = 175 - 0.093 (P/\text{MPa}),
\end{equation}
which is shown in \figref{fig:PTdiagram}, and an amplitude of\footnote{For values of
$\amp$ in the given range, the sum of squared deviations of the model from the data
varies by 5\% or less.}
\begin{equation}
    \amp = 2.3 \pm 0.4.
\end{equation}
As shown in the appendix, the amplitude $\amp$ is related to the coupling constant
$\beta$ as $\amp \simeq 2\beta$. To verify whether there is a microscopic manifestation
of weak crystallization theory, we have computed the normalized short-wavelength density
fluctuations, corresponding to the fluctuations of the order parameter $\psi$ from weak
crystallization theory, at atmospheric pressure as a function of temperature. The
quantity plotted in \figref{fig:qoft} is dimensionless but its magnitude is not defined
unambiguously. It is calculated by taking the expectation value over the Fourier
transform of the density operator which results in the form $\langle\exp(\mathrm{i}\, q
r) \exp(-\mathrm{i}\, q r)\rangle$ where $r$ is a particle position and $q$ is the
wavenumber. The wavenumber is chosen at the highest peak in the structure factor. The
computation is performed for an $N=8000$ particle system at a constant pressure of
0.1~MPa, and each point is averaged over 10~ns. The fluctuations would diverge in the
thermodynamic limit for a crystal, illustrating the broken symmetry. In the liquid phase,
the order parameter $\langle\psi\rangle = 0$. \ffigref{fig:qoft} shows the order
parameter fluctuations together with a fit of the form
\begin{equation}\label{eq:orderparameterfit}
    \langle\delta\psi\rangle^2 =  b \bigl[(T-\Ts)/\Ts\bigr]^{-1/2},
\end{equation}
where $b$ is a constant proportional to the coupling constant $\beta$ in \eqref{eq:gap}.
From the fit we obtain $b \simeq 0.24$ and $\Ts \simeq 181$~K, close to the value of
175~K found above. Another way to estimate $\Ts$ is by extrapolating the surface tension
between liquid and crystal as a function of supercooling, and finding the temperature at
which it becomes zero. An extrapolation of the surface tension values of Limmer and
Chandler\cite{limmer2012} yields $\Ts \simeq$ 170~K. The stability-limit temperature
$T\ts{s}$ lies far below the temperature of maximum crystallization rate determined by
Moore and Molinero,\cite{moore2011} which is 202~K at 0.1~MPa and signals the lowest
temperature at which liquid mW water can be equilibrated. This temperature difference
should not be interpreted as a disagreement, because the temperature of maximum crystal
growth is a kinetic quantity and does not correspond to a thermodynamic instability.

While the weak crystallization model provides a reasonable description of the mW data as
shown in \figref{fig:WC}, the amplitude $\amp \simeq 2\beta \simeq 2$ is not small
compared to unity, contradicting \eqref{eq:gap}, which is based on the assumption of
small fluctuation corrections in the theory. The fit based on weak crystallization theory
also implies that the density maximum at atmospheric pressure is caused by translational
short-wavelength fluctuations of density. This is unphysical because such fluctuations
should not have a significant effect far (100~K) from the stability limit of the liquid
state, where the density maximum is observed. Moreover, weak crystallization theory
predicts universal fluctuation-induced corrections to the regular behavior of
thermodynamic properties and is equally applicable to all metastable fluids, not only to
tetrahedral fluids that expand on freezing and exhibit a density maximum.

That the weak-crystallization power laws provide a good empirical description of water's
anomalies is not surprising. The properties of real supercooled water can also be
described quite well with purely empirical power laws diverging along a certain line
below the homogenous ice nucleation limit.\cite{speedy1976,kanno1979} The fitted line of
apparent singularities and the fitted values of the exponents are strongly correlated and
cannot be obtained independently. The larger the exponent value, the further away from
the homogenous nucleation this line is shifted. Adjustable amplitudes of the power laws
would increase the ambiguity even further.

However, we cannot ignore the fact that the fluctuations of the crystallization order
parameter indeed increase upon supercooling, as demonstrated in \figref{fig:qoft}. While
the relation between the amplitude of this effect and the coupling constant $\beta$ of
weak crystallization theory is unclear, it seems quite plausible that the fluctuations
also contribute to the thermodynamic anomalies.

\section{Conclusions}
By fitting properties with equations of state based on different underlying assumptions,
the consistency of such assumptions with thermodynamics can be tested. In this way we
have examined the origin of the anomalous behavior of the mW model in terms of a
two-state model and a model based on weak crystallization.

We have reproduced the anomalies of the mW model with a phenomenology based on a
non-ideal mixture of two different molecular configurations in liquid water. While an
ideal mixture of these states also generates enhancements in the response functions,
agreement with the mW data on the low-density fraction unambiguously requires a non-ideal
mixture. However, the nonideality is not strong enough to cause liquid--liquid phase
separation in the mW model, in the wide range of pressures and temperatures of this
study. An entropy-driven, athermal-solution-like nonideality gives a better description
of the mW data than an energy-driven, regular-solution-like nonideality. Incorporating
the formation of clusters into the equation of state results in a further improvement of
the description of the low-density fraction in mW water. The thermodynamic treatment of
the two-state model used here is independent of the details of the states considered.
Microscopically, however, there is no unique projection for defining these states from
the continuum distribution of configurations.

A description which uses the power laws of weak crystallization theory succeeds in
reproducing the calculated mW properties. The required-by-fit value of the coupling
constant appears to be about 2, which contradicts the assumption of the theory where this
constant is essentially a small parameter. Finally, weak crystallization theory is based
on the existence of the liquid-state stability limit and accounts for the growing
short-wave density fluctuations near that limit, which indeed are found in
simulations.\cite{moore2011,limmer2011} However, unlike the two-state approach, it cannot
account for the data on the fraction of four-coordinated molecules. The need to consider
fluctuation effects due to crystallization is consistent with previous work on the
modulation of ice interfaces in real water.\cite{limmer2012} This previous work
demonstrated that such fluctuations are important in describing how crystallization is
altered in confinement and suggests that such effects can be correctly accounted for
within simple Gaussian corrections. Reconciling the anomalous thermodynamics that are
well recovered by the two-state model, with the unavoidable effects of crystallization is
an important avenue to pursue in the future.

\begin{acknowledgments}
The authors have benefited from numerous interactions with Pablo Debenedetti (Princeton
University). Jan V. Sengers (University of Maryland, College Park) read the manuscript
and made useful comments. M.A.A. acknowledges discussions with Efim I. Kats (Landau
Institute, Russia) on weak crystallization theory. The research of V.H. and M.A.A. has
been supported by the Division of Chemistry of the U.S. National Science Foundation under
Grant No. CHE-1012052. D.T.L. acknowledges the Helios Solar Energy Research Center, which
is supported by the Director, Office of Science, Office of Basic Energy Sciences of the
U.S. Department of Energy under Contract No. DE-AC02-05CH11231. V.M. acknowledges support
by the National Science Foundation through awards CHE-1012651 and CHE-1125235 and the
Camille and Henry Dreyfus Foundation through a Teacher-Scholar Award.
\end{acknowledgments}

\section*{Appendix}
\subsection*{Expressions for thermodynamic properties in weak crystallization theory}
From derivatives of the chemical potential of \eqref{eq:gibbsWC}, we obtain
\begin{align}
    \Vh &=  \pypxd{\mh}{\Ph}{T}   =\frac{\amp a' \Th}{2\Ths^2\Delta^{1/2}} + \mr_{\Ph},\\
    \Sh &= -\pypxd{\mh}{\Th}{P}   =-\frac{\amp}{2\Ths\Delta^{1/2}} - \mr_{\Th},\\
    \Cph &= \Th\pypxd{\Sh}{\Th}{P} =
    \Th \left(\frac{\amp}{4\Ths^2\Delta^{3/2}} - \mr_{\Th\Th} \right),\\
    \kap &= -\frac{1}{\Vh}\pypxd{\Vh}{\Ph}{T} = \frac{1}{\Vh}
                    \left(\frac{\amp a'^2\Th^2}{4\Ths^4\Delta^{3/2}} - \frac{\amp a'^2\Th}{\Ths^3\Delta^{1/2}}
                    - \mr_{\Ph\Ph}\right),\\
    \alp &= \frac{1}{\Vh}\pypxd{\Vh}{\Th}{P} =\frac{1}{\Vh}
                    \left(-\frac{\amp a'\Th}{4\Ths^3\Delta^{3/2}} + \frac{\amp a'}{2\Ths^2\Delta^{1/2}}
                    + \mr_{\Th\Ph}\right).
\end{align}
where $\Vh = V/V_0$, $\Sh$, $\Cph$, $\kap$ and $\alp$ are the dimensionless volume,
entropy, isobaric heat capacity, isothermal compressibility and the thermal expansivity,
respectively. The subscripts of $\mr$ indicate a partial derivative of $\mr$ with respect
to the subscripted quantities.

\subsection*{Relation between the amplitude $\amp$ and the coupling constant $\beta$
in weak crystallization theory}
Consider the temperature-squared term in the dimensionless
chemical potential $\mh$,
\begin{equation}
    c \Th^2 = c \Ths^2 (\Delta + 1)^2 = c \Ths^2 (\Delta^2 + 2\Delta +1).
\end{equation}
Considering the $\Delta^2$ term, using \eqref{eq:gap} gives
\begin{align}
    \Delta^2    &= \Delta_0^2 + 2\beta\Delta_0\Delta^{-1/2} + \beta^2\Delta^{-1}\notag\\
                &= \Delta_0^2 + 2\beta\Delta^{1/2} - \beta^2\Delta^{-1}.
\end{align}
The last term can be neglected because according to the theory $\beta\ll1$. Since this is
not the case in our attempt to apply weak crystallization theory to the mW data, such a
description becomes purely empirical.

The first term on the right-hand side $\Delta_0^2$ is part of the regular part of the
chemical potential, and the second term $2\beta\Delta^{1/2}$ leads to the fluctuation
contribution $\delta\mh$ in the chemical potential of
\begin{equation}
    \delta\mh = 2c \Ths^2 \beta \Delta^{1/2}.
\end{equation}
Comparing this with \eqref{eq:gibbsWC} gives the relation between $\amp$ and $\beta$,
\begin{equation}\label{eq:a_lambda}
    \amp = 2c \Ths^2 \beta.
\end{equation}
For our fit to the mW data, $c\simeq 1$ and $\Ths\simeq 1$, so \eqref{eq:a_lambda} gives
\begin{equation}
    \amp \simeq 2\beta.
\end{equation}

\subsection*{Parameter values}
\begin{table}
\caption{\label{tab:twostate}Parameters for the two-state equation of state,
\eqref{eq:athermalsolution}}
\begin{ruledtabular}
\begin{tabular}{lD{.}{.}{12}lD{.}{.}{12}}
Parameter & \multicolumn{1}{c}{Value} & Parameter & \multicolumn{1}{c}{Value}\\\hline
$T_0$	&	203.07 \text{ K}	&	$c_{03}$	&	1.5125\times 10^{-4}	\\
$\rho_0$	&	1000.0 \text{ kg m$^{-3}$}	&	$c_{05}$	&	-1.5795\times 10^{-7}	\\
$\lambda$	&	2.6917	&	$c_{11}$	&	8.4408\times 10^{-2}	\\
$\omega_0$	&	1.6362	&	$c_{12}$	&	-3.9163\times 10^{-3}	\\
$\omega_1$	&	1.6777	&	$c_{13}$	&	1.2385\times 10^{-4}	\\
$\Ph_1$	&	2.3219	&	$c_{20}$	&	-1.7092\times 10^{0}	\\
$a$	&	0.039968	&	$c_{21}$	&	-4.3879\times 10^{-2}	\\
$c_{01}$	&	9.5440\times 10^{-1}	&	$c_{30}$	&	4.0687\times 10^{-1}	\\
$c_{02}$	&	-5.0517\times 10^{-3}	&	$c_{31}$	&	6.1937\times 10^{-2}	\\
\end{tabular}
\end{ruledtabular}
$\chi^2 = 0.82$
\end{table}

\begin{table}
\caption{\label{tab:twostatewithclustering}Parameters for the two-state equation,
\eqref{eq:athermalsolutionwithclustering}, with $N=6$}
\begin{ruledtabular}
\begin{tabular}{lD{.}{.}{12}lD{.}{.}{12}}
Parameter & \multicolumn{1}{c}{Value} & Parameter & \multicolumn{1}{c}{Value}\\\hline
$T_0$	&	203.07	&	$c_{03}$	&	2.8467\times 10^{-4}	\\
$\rho_0$ &	1000.0 \text{ kg m$^{-3}$}	&	$c_{05}$	&	-3.1836\times 10^{-7}	\\
$\lambda$	&	1.529	&	$c_{11}$	&	3.9710\times 10^{-2}	\\
$\omega_0$	&	0.19523	&	$c_{12}$	&	-6.4543\times 10^{-4}	\\
$\omega_1$	&	0.19856	&	$c_{13}$	&	4.6055\times 10^{-5}	\\
$\Ph_1$	&	1.1637	&	$c_{20}$	&	-1.8630\times 10^{0}	\\
$a$	&	0.039968	&	$c_{21}$	&	-3.2135\times 10^{-2}	\\
$c_{01}$	&	9.8530\times 10^{-1}	&	$c_{30}$	&	3.3420\times 10^{-1}	\\
$c_{02}$	&	-8.1978\times 10^{-3}	&	$c_{31}$	&	5.7351\times 10^{-2}	\\
\end{tabular}
\end{ruledtabular}
$\chi^2 = 0.72$
\end{table}

\begin{table}[t!]
\caption{\label{tab:WC}Parameters for the fit of weak crystallization theory,
\eqref{eq:gibbsWC}}
\begin{ruledtabular}
\begin{tabular}{lD{.}{.}{12}lD{.}{.}{12}}
Parameter & \multicolumn{1}{c}{Value} & Parameter & \multicolumn{1}{c}{Value}\\\hline
$\Tsz$	&	175.28 \text{ K}	&	$c_{11}$	&	4.2389\times 10^{-1}	\\
$M/V_0$	&	1000.0 \text{ kg m$^{-3}$}&	$c_{12}$	&	-3.3672\times 10^{-4}	\\
$\amp$	&	2.3254	&	$c_{13}$	&	-1.3649\times 10^{-4}	\\
$a'$	&	0.042825	&	$c_{20}$	&	3.4178\times 10^{-1}	\\
$c_{01}$	&	6.1915\times 10^{-1}	&	$c_{21}$	&	-2.3514\times 10^{-1}	\\
$c_{02}$	&	-8.2724\times 10^{-3}	&	$c_{30}$	&	-2.2560\times 10^{-1}	\\
$c_{03}$	&	3.6026\times 10^{-4}	&	$c_{31}$	&	4.9163\times 10^{-2}	\\
$c_{05}$	&	-6.6486\times 10^{-7}	
\end{tabular}
\end{ruledtabular}
$\chi^2 = 0.72$
\end{table}

Parameters for the two-state equation of state, \eqref{eq:athermalsolution}, are listed
in \tabref{tab:twostate}, and those for the fit of the equation with clusters of six
molecules, \eqref{eq:athermalsolutionwithclustering}, are given in
\tabref{tab:twostatewithclustering}. The value of the interaction parameter $\omega$
depends quadratically on the pressure,
\begin{equation}
    \omega(\Ph) = \omega_1 - (\omega_1 - \omega_0)\bigl[(\Ph-\Ph_1)/\Ph_1\bigr]^2,
\end{equation}
where $\omega_0$ is the value of $\omega$ at $\Ph=0$, $\omega_1$ is the maximum value of
$\omega$, and $\Ph_1$ is the dimensionless pressure at which the maximum occurs. The
parameter values for the fit of weak crystallization theory to the mW data are given in
\tabref{tab:WC}. All tables give $\chi^2$, the reduced sum of squared deviations of the
fit from the data.

\bibliography{vincentnew,supercooled}

\end{document}